\documentclass[aps,amsfonts,twocolumn,nofootinbib]{revtex4}
\usepackage{amsmath}
\usepackage{epsfig}
\usepackage{latexsym}
\usepackage{color}

\def\de{\partial}
\def\a{\alpha}
\def\b{\beta}
\def\g{\gamma}

\def\d{\delta}

\def\k{\kappa}
\def\m{\mu}
\def\n{\nu}
\def\r{\rho}

\def\th{\theta}
\def\z{\zeta}
\def\be{\begin{equation}}
 \def\ee{\end{equation}}
 \def\bea{\begin{eqnarray}}
 \def\eea{\end{eqnarray}}
 \def\a{\alpha}
 \def\b{\beta}
 \def\g{\gamma}
 \def\d{\delta}

\def\L{\Lambda}

\def\2{\frac{1}{2}}
\def\4{\frac{1}{4}}

\catcode`\@=11

\catcode`\@=12

\begin{document}

\title{String-Like BTZ  on Codimension-2 Braneworlds in the Thin Brane Limit}

\author{Bertha Cuadros-Melgar$^{1}$,
Eleftherios Papantonopoulos$^{2}$,  Minas Tsoukalas$^{2}$,
Vassilios Zamarias$^{2}$}

\affiliation{\vspace*{0.2cm} $^1$Departamento de F\'isica,
Universidad de Santiago de Chile, Casilla 307, Santiago, Chile}

\affiliation{$^2$Department of Physics, National Technical
University of Athens, GR~157~73~Athens, Greece}

\date{\today}

\begin{abstract}

We consider five-dimensional gravity with a Gauss-Bonnet term in
the bulk and an induced gravity term on a 2-brane of
codimension-2. We show that this system admits  BTZ black holes on
the 2-brane which are extended into the bulk with regular horizons.

\end{abstract}

\maketitle
%
%


%

%
%

%




%


%



A  growing interest in codimension-2 braneworlds, i.e a brane
embedded in a bulk with two extra dimensions, has recently
appeared. The most attractive feature of these models is that the
vacuum energy (tension) of the brane
 instead of curving the brane world-volume, merely induces a deficit angle in the
 bulk  around the brane \cite{Chen:2000at}. This property was used to solve the cosmological constant problem \cite{6d}.
However, soon it was realized \cite{Cline} that one can only find
nonsingular solutions if the brane stress tensor is proportional
to its induced metric. To obtain the Einstein equation on the
brane one has to introduce a cut-off (brane
thickness)\cite{Kanno:2004nr},  loosing the predictability of the
theory, or alternatively,
one can modify the gravitational action by including a
Gauss-Bonnet term~\cite{Bostock:2003cv} or a scalar curvature term
(induced gravity) on the brane~\cite{Papantonopoulos:2005ma}.

We still lack an understanding of time dependent
cosmological solutions in codimension-2 braneworlds. In the thin
brane limit, because the energy momentum tensors on the brane and
in the bulk are related,  we cannot get the standard cosmology on
the brane~\cite{Kofinas:2005py,Papantonopoulos:2005nw}. One then
has to regularize the codimension-2 branes by introducing some
thickness and then consider matter on them \cite{regular,tas}.
A cosmological evolution on the regularized branes requires an expanding brane
world-volume
and in general also a time evolving bulk.
An alternatively approach was followed in~\cite{Papantonopoulos:2007fk}
by considering a codimension-1 brane moving in the regularized
static background. The resulting cosmology, however, was unrealistic
having a negative Newton's constant (for a review
see~\cite{Papantonopoulos:2006uj}).

Moreover, the issue of localization of a black hole on the brane and its extension to the bulk is not fully understood.
In codimension-1 braneworlds, a first attempt was to consider
the black string extension in the bulk of a Schwarzschild metric
~\cite{Chamblin:1999by}.
Unfortunately, this string is unstable to classical linear
perturbations~\cite{BSINS} (for a review
see~\cite{Harmark:2007md}).
Further attempts deal with the Einstein equations projected on the brane, which include an unknown bulk dependent term, the Weyl tensor projection. Due to this reason the system is not closed, and some
assumptions have to be made either in the form of the metric or in
the Weyl term~\cite{BBH}.
The stability and
thermodynamics of these solutions were worked out in \cite{ACMPM}.

A lower dimensional version of a black hole living on a
(2+1)-dimensional braneworld was considered in~\cite{EHM} by
Emparan, Horowitz, and Myers. They based their analysis on the
so-called C-metric~\cite{Kinnersley:zw} modified by a cosmological
constant term. They found a BTZ black hole~\cite{Banados:1992wn}
on the brane which can be extended as a BTZ  string in a
four-dimensional AdS bulk. Their thermodynamical stability
analysis showed that the black string remains a stable
configuration when its transverse size is comparable to the
four-dimensional AdS radius, being destabilized by the
Gregory-Laflamme instability above that scale, breaking up to a
BTZ black hole on a 2-brane.

In the codimension-2 scenario, a six-dimensional black hole
 on a 3-brane was proposed in
~\cite{Kaloper:2006ek} and extended in~\cite{Kiley:2007wb} to
include also rotations. This is a generalization of the $4D$
Aryal, Ford, Vilenkin~\cite{Aryal:1986sz} black hole pierced by a
cosmic string adjusted to the codimension-2 branes, with a conical
structure in the bulk and deformations accommodating the deficit
angle. However, it is not clear how to realize these solutions in
the thin 3-brane limit.


In this work we study black holes on an infinitely thin conical
2-brane and their extension into the five-dimensional bulk.
We consider the following gravitational action in five dimensions
with a Gauss-Bonnet term~\cite{Charmousis:2002rc} in the bulk and an induced
three-dimensional curvature term on the brane
\begin{eqnarray}\label{AcGBIG}
S_{\rm grav}&=&\frac{M^{3}_{(5)}}{2}\Big{\{}\int d^5 x\sqrt{-
g^{(5)}}\Big{[} R^{(5)}
+\alpha \Big{(} R^{(5)2} \nonumber \\
&-& 4R^{(5)}_{MN}R^{(5)MN}+
R^{(5)}_{MNKL}R^{(5)MNKL}\Big{)}\Big{]} \nonumber\\&+& r^{2}_{c}
\int d^3x\sqrt{- g^{(3)}}\,R^{(3)}\frac{\delta(\rho)}{2\pi
b}\Big{\}}\nonumber
\\&+&\int d^5 x \mathcal{L}_{bulk}+\int d^3 x
\mathcal{L}_{brane}\frac{\delta(\rho)}{2\pi b}\,,\label{5daction}
\end{eqnarray}
where $\alpha\, (\geq0)$ is the GB coupling constant,
$r_c=M_{(3)}/M_{(5)}^3$ is the induced gravity ``cross-over"
scale,
$M_{(5)}$ is the five-dimensional Planck mass, and $M_{(3)}$ is the three-dimensional one.
 The above induced term has been
written in the particular coordinate system in which the metric is
\be
ds_5^2=g_{\m\n}(x,\rho)dx^\m
dx^\n+a^{2}(x,\rho)d\rho^2+b^2(x,\rho)d\th^2~,\label{5dmetric} \ee
where $g_{\mu\nu}(x,0)$ is the braneworld metric, and $x^{\mu}$
denotes three  dimensions, $\mu=0,1,2$, whereas $\rho,\th$ denote
the radial and angular coordinates of the two extra dimensions, and
we have assumed  an azimuthal symmetry in the system. Capital
$M$,~$N$ indices take values in the five-dimensional space.

The Einstein equations resulting from the variation of the
action~(\ref{5daction}) are

\bea
 G^{(5)N}_M &+& r_c^2
G^{(3)\n}_\m g_M^\m g^N_\n {\d(\rho) \over 2 \pi b}-\alpha
H_{M}^{N}\nonumber \\ &=&\frac{1}{M^{3}_{(5)}}
\left[T^{(B)N}_M+T^{(br)\n}_\m g_M^{\,\m} g^N_\n {\d(\rho) \over 2 \pi
b}\right]~, \label{einsequat} \eea where $H_M^N$ is the
Gauss-Bonnet contribution to the bulk
equations~\cite{Bostock:2003cv}.
To obtain the braneworld equations we expand the metric around the
brane as $ b(x,\rho)=\beta(x)\rho+O(\rho^{2})~.$

 At the boundary
of the internal two-dimensional space where the 2-brane is
situated, the function $b$ behaves as $b^{~\prime}(x,0)=\beta(x)$,
where a prime denotes derivative with respect to $\rho$. We also
demand that the space in the vicinity of the conical singularity
is regular, which imposes the supplementary conditions $\de_\m
\b=0$ and
$\partial_{\rho}g_{\mu\nu}(x,0)=0$~\cite{Bostock:2003cv}.

The extrinsic curvature in the particular gauge $g_{\rho \rho}=1$
that we consider is given by $K_{\m\n}=g'_{\m\n}$.
 We will now use the fact that the second derivatives of the metric
functions contain $\d$-function singularities at the position of
the brane. The nature of the singularity then gives the following
relations \cite{Bostock:2003cv}
\bea
{b'' \over b}&=&-(1-b'){\d(\rho) \over b}+ {\rm non-singular~terms}~,\\
{K'_{\m\n} \over b}&=&K_{\m\n}{\d(\rho) \over b}+ {\rm
non-singular~terms}~. \eea

From the above singularity expressions and using the Gauss-Codacci
equations, we can  match the singular parts of the Einstein
equations (\ref{einsequat}) and get the following ``boundary"
Einstein equations
 \be G^{(3)}_{\m\n}={1 \over M_{(5)}^3
(r_c^2+8\pi (1-\b)\a)}T^{(br)}_{\m\n}+{2\pi (1-\b) \over
r_c^2+8\pi (1-\b)\a}g_{\m\n} \label{einsteincomb} \ee


We assume that there is a  (2+1) black hole on the brane. The
brane metric is \be
ds_{3}^{2}=\left(-n(r)^{2}dt^{2}+n(r)^{-2}dr^{2}+r^{2}d\phi^{2}\right)~,
\label{3dmetric}\ee where $0\leq r< \infty$ is the radial
coordinate, $\phi$ has
 the usual periodicity $(0,2\pi)$ and $l$ is the
length scale of the $AdS_{3}$
 space.
 We will look for string-like solutions of the
 Einstein equations~(\ref{einsequat}) using the
 five-dimensional metric~(\ref{5dmetric}) in the form
\bea
ds_5^2&=&f^{2}(\rho)\left(-n(r)^{2}dt^{2}+n(r)^{-2}dr^{2}+r^{2}
d\phi^{2}\right)\nonumber
\\ &+&a^{2}(r,\rho)d\rho^2+b^2(r,\rho)d\th^2~.\label{5smetric} \eea

Since the space outside the
 conical singularity is regular, the warp function $ f(\rho) $ must
  also be regular
 everywhere. We assume that there is only a cosmological constant $\Lambda_{5}$
in the bulk, and we take $a(r,\rho)=1$. Then
 combining the $(rr,\phi \phi)$ Einstein equations
  we get
 \be
 \left(\dot{n}^{2}+n \ddot{n}-\frac{n \dot{n}}{r}\right)\left(1-4\alpha \frac{b''}{b}\right)=0~,\label{17}
 \ee
 while a combination of the $(\rho\rho, \theta \theta)$ equations
  gives
 \be
 \left(f''-\frac{f'b'}{b}\right)\left[3-4\frac{\alpha}{f^{2}}\left(\dot{n}^{2}+n
  \ddot{n}+2\frac{n \dot{n}}{r}+3f'^{2}
 \right)\right]=0\label{18}~,
 \ee
 where a  dot denotes derivatives with respect to $r$.
We will consider first 
$\dot{n}^{2}+n \ddot{n}-\frac{n \dot{n}}{r}=0~,$
which has as a solution  the simplest BTZ black hole without
charge or angular momentum~\cite{Banados:1992wn}
 \be n^{2}(r)=-M+\frac{r^{2}}{l^{2}}~.\label{btz} \ee
Then equation (\ref{18}) becomes \be
 \left(f''-\frac{f'b'}{b}\right)\left[1-4\frac{\alpha}{f^{2}}\left(\frac{1}{l^{2}}+f'^{2}
 \right)\right]=0\label{18a}~.
 \ee
From the above equation we have two cases. The first case is
$f'^{2}-\frac{f^{2}}{4\alpha}+\frac{1}{l^{2}}=0~, \label{1stcase}$
which has the following solution
 \be
f_{1}(\rho)=C_{1}\,e^{\frac{\rho}{2\sqrt{\alpha}}} +
C_{2}\,e^{\frac{-\rho}{2\sqrt{\alpha}}}~,\label{solu1}
 \ee where $
C_{1}$ and $C_{2}$ are integration constants and satisfy the
relation $C_{1}\,C_{2}=\a/l^2$. The function $f(\rho)$ is regular
and if we require that on the position of the brane the boundary
condition is $f^{2}(\rho=0)=1$, then  the integration constants
can be expressed in terms of $\alpha$ and $l$
 \be
C_{1}=\pm \frac{1 + \varepsilon
\sqrt{1-4\frac{\a}{l^2}}}{2}~,\,\,\,\, C_{2}=\pm \frac{1 -
\varepsilon \sqrt{1-4\frac{\a}{l^2}}}{2}~~,
 \ee
where $\varepsilon=\pm 1$ independently of the $\pm$ sign in
$C_{1}$ and $C_{2}$. If we impose also the condition
 $\partial_{\rho} g_{\mu\nu}(x,0)=0$ we obtain $C_1=C_2=1/2$ and the solution
(\ref{solu1}) simplifies to $f_{1}(\rho)=cosh(\rho/2\sqrt{a})$.
Substituting the above solutions back to the five-dimensional
equations  we get a fine-tuned relation between $\alpha$ and
$\Lambda_{5}$ \be \Lambda_{5}=-\frac{3}{4 \alpha}~.\label{tuning}
 \ee
 Because of
the positivity of $\alpha$ the five-dimensional bulk space is
Anti-de-Sitter.
The choices we made in solving (\ref{17}) and (\ref{18})
determined only the functions $n(r)$ and $f(\rho)$ and  although
they solve equations (\ref{einsequat}) they leave $b(\r)$
undermined making the bulk metric
arbitrary~\cite{Charmousis:2002rc}.

The second case is to consider \be f''-\frac{f' b '}{b}=0
\Rightarrow b(\r)=b_0\,f'(\r)~.\label{25a} \ee Then, back into the
bulk equation we notice that the $(\rho \rho)$  equation can only
be solved when $\Lambda$ takes the same value as in the first case
given in (\ref{tuning}), thus we have
 \be
 \left(1-4\,\a\,\frac{f''}{f}\right)\left[1-4\frac{\alpha}{f^{2}}\left(\frac{1}{l^{2}}+f'^{2}
 \right)\right]=0\label{18b}~.
 \ee
from which we have two subcases. The first subcase is
$\left(1-4\,\a\,\frac{f''}{f}\right)=0$, and with
(\ref{tuning}) it gives the same solution for $f(\r)$ as in equation
(\ref{solu1})
where $C_{1}$ and $C_{2}$ are integration constants. Imposing
again the boundary condition $f^{2}(\rho=0)=1$ and the fact that
at the position of the brane $b(\r=0)=0$ and $b'(\r=0)=\b$ we get
$C_{1}=C_{2}=\pm \frac{1}{2}$ and in (\ref{25a}) $b_0=4\,\a\,\b$.
Therefore, for this case $f(\r)$ and $b(\r)$ can be written as
\be f_{2}(\r)=\pm \cosh\left(\frac{\r}{2\,\sqrt{\a}}\right),\,
b_{2}(\r)=\pm
2\,\b\,\sqrt{\a}\,\sinh\left(\frac{\r}{2\,\sqrt{\a}}\right).
\label{bsolu2a} \ee One can check that the above solution is
consistent with all bulk equations.

For the second subcase we get as in the first case the same
solution for $f(\r)$ (Eq.(\ref{solu1})) with
$C_{1}\,C_{2}=\a/l^2$ but the function $b(\r)$ is given by
$b(\r)=b_0\,f'(\r)$. Imposing again the boundary conditions for
$f^{2}(\rho)$ and $b(\r)$ we get $b_0=2\,\a\,\b$ and
$C_{1}=C_{2}=1/2$. Then $f(\r)$ and $b(\r)$ are given by
(\ref{bsolu2a}), relation (\ref{tuning}) still holds, and we get
an extra constraint $ l^2=4\,\a.$

These solutions extent the
 BTZ black hole on the brane into the
bulk. Calculating the curvature invariants we find no $r=0$
curvature singularity for the BTZ string-like solution  as
expected.
The warp function $f^{2}(\rho)$ gives the shape of the horizon of
the BTZ string-like solution. The size of the horizon is defined
by the scale $\sqrt{\alpha}$. Using the fine-tuned
relation~(\ref{tuning}) and the relation $\Lambda_{5}=-6/L^{2}$,
 this
scale can be fine-tuned to the length scale $L$  of the
five-dimensional AdS space. Then,  the
   warp function $f(\rho)$ is finite at the boundary of the AdS space, and depending
   on the integration constants of the various cases, it gives the  shape of a 'throat' to
    the horizon.

There is also a constant solution for $f(\r)$ which we show in
Table \ref{table1} (with $\g=\sqrt{\frac{l^2-4\a}{2}}$ and
$\Lambda=-3/l^{2}$).


We have also studied more general solutions without the
restriction that $n$ is chosen as BTZ black hole, which means that
in equation (\ref{17}) we chose 
$1-4 \alpha \frac{b''}{b}=0~$,
from which $b(\r)$ is obtained as
\be b_{5}(\rho)=b_0
\left(C_{1}~e^{\rho/2\sqrt{\alpha}}+C_{2}~e^{-\rho/2\sqrt{\alpha}}\right)~.
\ee

The first case is to consider from (\ref{18}) the relation
(\ref{25a}).
Then we get
\be
f_{5}(\rho)=f_0\left(C_{1}\,e^{\frac{\rho}{2\sqrt{\alpha}}}
-
C_{2}\,e^{\frac{-\rho}{2\sqrt{\alpha}}}\right)+C_{3}~,\label{solu4}
 \ee where $
C_{1}$, $C_{2}$, and $C_{3}$ are integration constants and the
 bulk equations  gives $C_{3}=0$ and the relation~(\ref{tuning}).
  Imposing again the boundary conditions for
$f^{2}(\rho)$ and $b(\r)$  we get $C_{1}=C_{2}=\pm \frac{1}{2}~$,
$f_0=1$ and $b_0=2\,\b\,\sqrt{\a}$ we get $f(\r)$ and $b(\r)$ as
in (\ref{bsolu2a}). The function $n(r)$ remains undetermined
connected with the brane matter
$T_{1}^{1}=T_{2}^{2}=nn'/r-\Lambda_{3}$,
$T_{3}^{3}=n'^2+nn''-\Lambda_{3}$ from (\ref{einsteincomb}). For
the second case we analyse from (\ref{18}) the term
$3\left(f^{2} -4\a\,f'^2\right)-4\a \left(\dot{n}^{2}+n
\ddot{n}+2\frac{n \dot{n}}{r}\right
 )=0~.$
The first term is a function of $\r$ while the second one is a
function of $r$. Therefore, each term should be, in general, equal
to a constant $\k$. We then have \be 3
\left(f^2-4\,\a\,f'^2\right)=\k~, \,\,
4\a\left(2\frac{n\dot{n}}{r}+n\ddot{n}+\dot{n}^2\right)=\k~, \ee
which give \bea
f_{6}(\r)&=&C_3\,e^{\frac{\r}{2\sqrt{\a}}}+\frac{\k}{12\,C_3}\,e^{\frac{-\r}{2\sqrt{\a}}}~, \label{fsolu5}\\
n(r)&=&\sqrt{C_5+\frac{\k}{12\,\a}\,r^2+\frac{C_6}{r}}~.
\label{nsolu5} \eea Imposing that $f^{2}(\rho=0)=1$ we get for the
function $f(\r)$ as in the first case of the BTZ solution
\be
f_{6}(\rho)=C_{3}\,e^{\frac{\rho}{2\sqrt{\alpha}}} +
C_{4}\,e^{\frac{-\rho}{2\sqrt{\alpha}}}~,\label{fsolu5final} \ee
where
 \be
C_{3}=\pm \frac{1 +\varepsilon \sqrt{1-\frac{\k}{3}}}{2}~,\,\,\,
C_{4}=\pm \frac{1 -\varepsilon \sqrt{1-\frac{\k}{3}}}{2}~~.
 \ee
Moreover, imposing the boundary conditions for $b(\r)$ its
solution is given by equation (\ref{bsolu2a}). If we impose also
the condition
 $\partial_{\rho} g_{\mu\nu}(x,0)=0$ then $\kappa=3$ and the solution
(\ref{fsolu5final}) simplifies to
$f_{6}(\rho)=cosh(\rho/2\sqrt{a})$. These solutions represent
BTZ-corrected black string with the usual $r=0$ curvature
singularity.
 If we redefine $C_5=-M$, $C_6=-\z$, we use
$l^2=4 \a$, we get the BTZ black hole solution with a short
distance correction term which corresponds to the BTZ conformally
coupled to a scalar field~\cite{Zanelli1996}
 \be
n(r)=\sqrt{-M+\frac{r^2}{l^2}-\frac{\z}{r}}~. \label{nsolu5final}
\ee
 Substituting the above solutions into  the $(\rho \rho)$ bulk equation we get  (\ref{tuning}).
Using the relation $\L_5=-6/L^2$ we get the  fine-tuned relation
$L^2=2\,l^2~.$ There is also a constant solution for $f(\r)$ with
$\L_5=-\frac{1}{4\a}$.
We summarize our results in the following table.
\begin{table}[here]
\begin{tabular}{|c|c|c|c|}
  \hline
  $n(r)$ & $f(\r)$ & $b(\r)$ \\
  \hline
   & $\cosh\left(\frac{\r}{2\,\sqrt{\a}}\right)$ & $\forall b(\r)$  \\
  BTZ & $\cosh\left(\frac{\r}{2\,\sqrt{\a}}\right)$ & $2\,\b\,\sqrt{\a}\,\sinh\left(\frac{\r}{2\,\sqrt{\a}}\right)$  \\
   & $\pm 1$ & $\g\,\sinh\left(\g^{-1}\,\r\right)$  \\
\hline
  $\forall n(r)$ & $\cosh\left(\frac{\r}{2\,\sqrt{\a}}\right)$ & $2\,\b\,\sqrt{\a}\,\sinh\left(\frac{\r}{2\,\sqrt{\a}}\right)$
   \\
\hline
  corrected & $\cosh\left(\frac{\r}{2\,\sqrt{\a}}\right)$ &
$2\,\b\,\sqrt{\a}\,\sinh\left(\frac{\r}{2\,\sqrt{\a}}\right)$  \\
 BTZ & $\pm 1$ & $2\,\b\,\sqrt{\a}\,\sinh\left(\frac{\r}{2\,\sqrt{\a}}\right)$
   \\
  \hline
\end{tabular}\\
\caption{Results}\label{table1}
\end{table}


To complete our solution with the introduction of the brane we
must  solve the corresponding junction conditions given by the
Einstein equations on the brane (\ref{einsteincomb}) using the
induced metric on the brane given by (\ref{3dmetric}). For the
case when $n(r)$ corresponds to the BTZ black hole (\ref{btz}),
and the brane cosmological constant is given by
$\Lambda_{3}=-1/l^{2}$, we found that the energy-momentum tensor
is null. Therefore, the BTZ black hole is localized on the brane
in vacuum.

When $n(r)$ is of the form given in (\ref{nsolu5final}),  we found
the following traceless energy-momentum tensor \be T_\alpha ^\beta
=  \hbox{diag } \left(
\frac{\zeta}{2r^3},\frac{\zeta}{2r^3},-\frac{\zeta}{r^3} \right)\,
,\label{braneEnerMom} \ee which is conserved on the brane,
$\bigtriangledown_\beta T_\alpha ^\beta =0$~\cite{Kofinas:2005a}.
The conformally coupled scalar field to the BTZ black hole does
not introduce an independent conserved charge, it only modifies
the energy-momentum tensor of the three-dimensional Einstein
equations. If we consider the energy-momentum tensor
in~\cite{Zanelli1996} necessary to sustain such solution, and we
take the  limit  $r/l<<1$,  we get the unexpected result that it
reduces to (\ref{braneEnerMom}) which is necessary to localize
this black hole on the conical 2-brane. A way to understand this
result is that because in this limit $r$ is very small, the black
hole will be localized around the conical singularity and
therefore, any matter will take a distributional form around this
singularity.
 Note also, that this solution is a result of the presence
of the Gauss-Bonnet term in the bulk. If we switch off the
Gauss-Bonnet coupling, then from relations (\ref{17}) and
(\ref{18}) it can be seen that only the BTZ black hole is a
solution.

\textit{In conclusion } we discussed black holes  on an infinitely
thin 2-brane of codimension-2 and their extension into a
five-dimensional AdS bulk. To have a three-dimensional gravity on
the brane we introduced a five-dimensional Gauss-Bonnet term in
the bulk and an induced gravity term on the 2-brane. We showed
that this system admits  (2+1) BTZ black holes solutions and their
short distance extension  on the 2-brane, while in the bulk these
solutions describe BTZ-like strings. Consistency of the
five-dimensional bulk equations requires a fine-tuned relation
between the Gauss-Bonnet coupling constant and the length of the
five-dimensional AdS space. The use of this fine-tuning gives to
the non-singular horizon the shape of a throat up to the boundary
of the AdS space.

We did not allow more severe singularities than conical. This
assumption has fixed the deficit angle to a constant value. It is
interesting to investigate how our solutions are modified in the
presence of a variable deficit angle.
Also, we did not discuss the thermodynamics and the stability
issue
 of our solutions. We expect, however, similar
 behaviour of our solutions as the one found in four dimensions~\cite{EHM}.
 These issues are under study, and they will be
 reported elsewhere.

Of course, the important issue is if this analysis can be applied
to a conical 3-brane. In our case the conical-like metric of the
BTZ black hole helped us to localize it on the brane and further
extent it into the bulk. A possible clue could be the BTZ short
distance corrected solution (\ref{nsolu5final}). From a
four-dimensional point of view it is a topological black hole in
AdS space. It would be interesting to investigate the possibility
of localization and further extension in the bulk of black holes
with symmetries other than spherical.

\textit{Acknowledgments} We thank C. Charmousis and F. M\'endez for
discussions.
 The work of B.C-M. is supported by Fondo Nacional
de Desarrollo Cient\'{i}fico y Tecnol\'ogico (FONDECYT), Chile,
under grant 3070009. The work of E.P., M.T., and V.Z. is partially
supported by the NTUA research program PEVE07 and by the European
Union through the Marie Curie Research and Training Network
UniverseNet (MRTN-CT-2006-035863).

\end{document}